\documentstyle[12pt,fleqn]{article}
\def\be{\begin{equation}}
\def\ee{\end{equation}}
\def\bea{\begin{eqnarray}}
\def\eea{\end{eqnarray}}
\def\l{\label}

\def\e{\epsilon}
\def\G{\Gamma}

\begin{document}
\begin{titlepage}


\vspace{.35in}

\begin{center}
\Large
{\bf Tilting the Primordial Power Spectrum with Bulk Viscosity}

\vspace{.3in}

\normalsize

\large{James E. Lidsey$^1$}

\normalsize
\vspace{.7cm}

{\em NASA/Fermilab Astrophysics Center, \\ Fermi National Accelerator
Laboratory, Batavia, IL~~ 60510. U.S.A.}

\end{center}

\vspace{.7in}

\baselineskip=24pt
\begin{quote}
\normalsize
\hspace*{2em}
Within the context of the cold dark matter model, current observations
suggest that inflationary models which generate a tilted primordial
power spectrum with negligible gravitational waves provide the most
promising mechanism for explaining large scale clustering. The general
form of the inflationary potential which produces such a spectrum is a
hyperbolic function and is interpreted physically as a bulk viscous
stress contribution to the energy-momentum of a perfect baryotropic
fluid. This is equivalent to expanding the equation of state as a
truncated Taylor series. Particle physics models which lead to such a
potential are discussed.

\vspace*{12pt}


\vspace*{12pt}
\noindent
\small $^1$email- jim@fnas09.fnal.gov

\vspace{1cm}
\normalsize

To Appear {\em Mon. Not. R. Astro. Soc}

\end{quote}

 \normalsize
\end{titlepage}


\section{Introduction}

\setcounter{equation}{0}

The origin and evolution of large scale structure is one of the most
important problems in cosmology today. It is widely accepted that the
growth of small fluctuations by gravitational instability leads to
structure formation. The inflationary paradigm, whilst providing
possible solutions to a number of other problems associated with the
hot big bang model, also produces a Gaussian, adiabatic fluctuation
spectrum which is nearly, though not exactly, scale-invariant (Guth
1981; Albrecht \& Steinhardt 1982; Linde 1982; Olive 1990; Liddle \&
Lyth 1993).  Based on this prediction, the standard Cold Dark Matter
(CDM) model of galaxy formation employs the flat, Harrison-Zel'dovich
spectrum as an input parameter (Efstathiou 1990). The CDM model
successfully accounts for small ($\le 10h^{-1}$ Mpc) and intermediate
($10h^{-1}$ Mpc - $100h^{-1}$ Mpc) scale observations, if one
introduces a bias in the distribution of luminous to dark matter
(Davies et al. 1985).{\footnote{The current value of the expansion
rate is $H_0 = 100h$ km ${\rm s}^{-1}$ ${\rm Mpc}^{-1}$, where $0.4\le
h \le 1$.}}

However, standard CDM has come under severe pressure from a number of
recent observations (for a detailed review see Liddle \& Lyth
1993). In particular, the APM angular galaxy-galaxy correlation
function (Maddox et al. 1991) and the {\em IRAS} QDOT redshift survey
(Efstathiou et al. 1991) indicate that there exists more large scale
structure than that predicted by standard CDM. One possible resolution
to this problem is to consider {\em tilted} CDM models. Here the
primordial power spectrum is assumed to be of the form $P(k) \propto
A_S^2(k)k \propto k^n$, where $k$ is the comoving wavenumber of the
Fourier expansion of the perturbation, $A_S$ is the amplitude of the
quantum fluctuation when it crosses the Hubble radius during the
matter- or radiation-dominated eras and $n$ is the power
spectrum. Other possibilities involve the addition of a cosmological
constant or a hot dark matter component (Liddle \& Lyth 1993).

Inflation also produces a spectrum of gravitational wave (tensor)
perturbations, whose amplitude may or may not be comparable to that of
the scalar fluctuations.  In this paper we shall concentrate on models
which lead to tilted power spectra with a negligible gravitational
wave component.  There exists a wide range of observational
constraints on the tilt arising from large angle ($\theta \ge 3^o$)
microwave background anisotropies (Smoot et al. 1992), galaxy
clustering (Maddox et al. 1990; Efstathiou et al. 1990), peculiar
velocity flows (Bertschinger \& Dekel 1989; Dekel, Bertschinger \&
Faber 1990; Bertschinger et al. 1990), high redshift quasars
(Efstathiou \& Rees 1988) and the red shift of structure formation
(Adams et al. 1993). When combined together these observations
strongly limit the allowed value of $n$. It has been shown that tilted
CDM can not fit all of the current data simultaneously (Adams et
al. 1993; Liddle \& Lyth 1993). For inflationary models in which
gravitational wave production is negligible, a lower limit of $n>0.7$
is partially consistent with the COBE 2-sigma upper limit and the bulk
flow data, if the clustering and pairwise velocity data are
ignored. In models where the gravitational wave contribution to the
microwave background anisotropy is important, however, this limit is
strengthened to $n>0.8 4$. This is clearly inconsistent with the APM
galaxy correlation function, which indicates that $0.3<n<0.7$ provides
a good fit to the excess clustering data.

On the other hand, there is growing observational evidence for a
departure from a pure power law at a scale $\lambda \approx 150 \pm 50
h^{-1}$ Mpc (Einasto et al. 1993). The power spectrum of clusters of
galaxies has spectral index $-2 \le n \le -1$ on intermediate scales,
whilst it is consistent with $n=1$ on large scales (Peacock 1991).

In short, the current status of the observations is far from
conclusive and it is therefore important to consider all the
theoretical options available. In this paper we investigate the
general circumstances in which a tilted power law scalar spectrum and
a negligible gravitational wave amplitude arise in inflationary
models. In Section 2, we summarize the details of a powerful framework
which allows the general form of the inflaton potential to be derived
in a straightforward manner. The form of such a potential is shown to
be a hyperbolic secant function in Section 3. In Section 4, it is
further shown that such a potential arises when a bulk viscous stress
is added to the energy-momentum tensor of a perfect baryotropic
fluid. It is illustrated how a number of plausible particle physics
models, such as the quantum creation of fundamental strings (Turok
1988) and $N=2$ supergravity (Salam \& Sezgin 1984), lead to a
potential of this form.

\vspace{.7cm}

\section{Inflaton dynamics and primordial fluctuations}

\vspace{.7cm}

Inflation proceeds if the potential energy of a scalar field dominates
the energy-momentum tensor of the universe, since this leads to a
violation of the strong energy condition (Hawking \& Ellis 1973).  One
can model self-interacting scalar fields in terms of perfect fluids,
and vice-versa. Within the context of the isotropic and homogeneous
Friedmann-Robertson-Walker (FRW) cosmologies, the complete history of
the universe is determined by three independent equations. The
Friedmann equation
\be
\l{1}
H^2 = \frac{\rho}{3}- \frac{k}{a^2}
\ee
is the first integral of the Raychaudhuri equation (Raychaudhuri 1955)
and describes the conservation of energy.  Here, $\rho(t)$ represents
the total energy density of the universe, $k = \{-1,0,+1\}$ for open,
flat, and closed spatial sections respectively, $a(t)$ is the scale
factor, and $H \equiv \dot{a}/a$ is the expansion rate, where a dot
denotes differentiation with respect to cosmic time $t$. Units are
chosen such that $\hbar=c=8\pi G =1$. For a matter content with
pressure $p(t)$, the Bianchi identity (equivalent to the local
conservation of energy-momentum) is
\be
\l{2}
\dot{\rho}+3H(\rho +p) =0,
\ee
and, in principle, the general solution to these equations follows
once the equation of state
\be
\l{3}
p=p(\rho)
\ee
is specified. This equation describes the particle physics sector of the model.

However, analytical solutions have only been found for a limited
number of specific examples, such as the special case $p/\rho={\rm
constant}$ (Barrow 1990). For an arbitrary equation of state which
satisfies the dominant energy condition $(\rho -p \ge 0)$, it proves
convenient to redefine the sum and difference of $\rho$ and $p$ in
terms of the new functions
\be
\l{4a}
{\dot{\phi}}^2 \equiv \rho +p \qquad \Longleftrightarrow \qquad \phi(t) =
\int^t dt' \sqrt{\rho (t') + p(t')}
\ee
\be
\l{4b}
2V \equiv \rho - p.
\ee
We may then rewrite equations (\ref{1}) and (\ref{2}) as
\bea
\l{5a}
H^2 = \frac{1}{3} \left( \frac{1}{2} {\dot{\phi}}^2+V \right) - \frac{k}{a^2}
\\
\l{5b}
\ddot{\phi}+3H\dot{\phi} + \frac{dV}{d\phi}=0,
\eea
which are the Einstein field equations for a minimally coupled scalar
field, $\phi$. Specifying the equation of state now becomes a question
of choosing an appropriate functional form for the potential
$V(\phi)$, and vice-versa. For example, an exponential potential is
equivalent to $p/\rho = {\rm constant}$ when $k=0$.

Recent advances in the treatment of equations (\ref{5a}) and
(\ref{5b}) have been made by viewing the scalar field as the effective
dynamical variable of the system (Muslimov 1990; Salopek \& Bond 1990,
1991; Lidsey 1991, 1993). From the definition $\rho \equiv
{\dot{\phi}}^2/2+V$, the time dependence can be eliminated by
rewriting the scalar field equation (\ref{5b}) as
\be
\l{6}
{\rho}' = -3H\dot{\phi}, \qquad \dot{\phi} \ne 0,
\ee
where a prime denotes differentiation with respect to $\phi$. This is
consistent if $\phi$ does not oscillate (i.e. $\dot{\phi}$ does not
pass through zero). It follows that $6H^2 =-\rho'X'/X$, where $X(\phi)
\equiv a^2(\phi)$, and the Friedmann equation becomes
\be
\l{7}
\rho'X'+2\rho X=6k.
\ee
The potential can be found immediately from the expression
\be
\l{8}
V(\phi)=\rho(\phi)-\frac{1}{18} \frac{(\rho')^2}{H^2(\phi)}
\ee
once the forms of $\rho(\phi)$ and $X(\phi)$ are known.

 When $k=0$, $3H^2=\rho$, and these field equations take the
particularly simple form
\be
\l{9}
2H'a' = -Ha, \qquad 2H'=-\dot{\phi},
\ee
thereby allowing the general solution, $a(\phi)$, to be expressed in
terms of quadratures with respect to $\phi$ (Salopek \& Bond 1991;
Lidsey 1993). The expression for the potential reduces to an
Hamilton-Jacobi differential equation in $H(\phi)$ of the form
\be
\l{10}
V(\phi) = 3H^2(\phi)-2(H')^2.
\ee

This framework is useful for determining the primordial power
spectrum.  During inflation short-wavelength quantum fluctuations in
the inflaton and graviton fields are redshifted beyond the Hubble
radius when the comoving wavenumber of the perturbation satisfies
$k=aH$. Once outside, their amplitude remains frozen until they
re-enter the Hubble radius during the radiation-dominated or
matter-dominated eras. The density perturbation (scalar) spectrum is
given by (Copeland et al. 1993)
\be
\l{3.1}
A_S(k) = \left( \frac{mH^2}{4\pi^{3/2}|\dot{\phi}|} \right)_{k=aH} = \left(
\frac{mH^2}{8\pi^{3/2}|H'|} \right)_{k=aH}
\ee
where the quantities on the right are calculated when the fluctuation
first crosses the Hubble radius. In the uniform Hubble constant gauge
the constant $m=4$ if the perturbation re-enters during the
radiation-dominated era, whereas $m=2/5$ if the perturbation re-enters
during matter domination.

The gravitational wave (tensor) spectrum is calculated in the
transverse-traceless gauge, where $+$ and $\times$ denote the two
independent polarization states of the metric perturbation. The
classical amplitude of the fluctuation satisfies the massless
Klein-Gordon equation (Grishchuk 1974, 1977; White 1992), so the
graviton can be viewed as a massless, minimally coupled scalar field,
which has two degrees of freedom $\psi_{+, \times}$. The spectrum of
tensor perturbations is then given by (Abbott \& Wise 1984)
\be
\l{3.2}
A_G(k)= \left( \frac{H}{4\pi^{3/2}} \right)_{k=aH}
\ee

One may define  two parameters (Copeland et al. 1993)
\be
\l{3.3}
\e \equiv 3 \frac{\dot{\phi}^2 /2}{V+\dot{\phi}^2 /2} = 2\left( \frac{H'}{H}
\right)^2
\ee
and
\be
\l{3.4}
\eta \equiv \frac{\ddot{\phi}}{H\dot{\phi}} = 2 \frac{H''}{H}.
\ee
 Up to a numerical factor, $\e$ measures the relative contribution of
the field's kinetic energy to its total energy density and may be
referred to as the {\em energy parameter}.  The quantity $\eta$
measures the ratio of the field's acceleration relative to the
friction contribution. We refer to it as the {\em friction parameter}.

These parameters may be recast in an alternative form by defining new variables
\be
\l{3.5}
x \equiv H^2, \qquad y \equiv (H')^2.
\ee
It follows from the identity $2H'' = d(H')^2/dH$ that
\be
\l{3.6}
\e = \frac{2y}{x}, \qquad \eta = 2 \frac{dy}{dx}.
\ee
Inflation proceeds in the region of parameter space for which $\e <1$
and the coasting solution, or Milne universe, corresponds to $\e =1$
({\em i.e.} $2y=x$). It is interesting to note that the friction
parameter does not directly determine whether inflation occurs. The
`slow-roll' approximation is valid when $\{\e, |\eta| \} \ll 1$.

One may also write the (scale-dependent) spectral indices of the
scalar and tensor fluctuations in terms of these two quantities.  It
is easy to show that
\bea
\l{3.7}
n-1  \equiv \frac{d \ln [A_S^2(k)]}{d \ln k} = 2  \left( \frac{2\e_*
-\eta_*}{\e_* -1} \right) \\
\l{3.8}
n_G \equiv \frac{d \ln [A_G^2(k)]}{d \ln k} = \frac{2 \e_*}{\e_* -1},
\eea
where $*$ indicates that $\e$ and $\eta$ should be evaluated when a
particular scale first crosses the Hubble radius. The flat
Harrison-Zel'dovich spectrum is equivalent to $n=1$.

It follows from the definitions of $A_S$ and $A_G$ that
\be
\l{3.9}
\frac{A_G}{A_S} =  \frac{\sqrt{2 \e}}{m}.
\ee
It is often stated that inflation leads to a Harrison-Zel'dovich
scalar spectrum with a negligible gravitational wave
contribution. However such a conclusion follows because the slow-roll
approximation is assumed {\em a priori}.  Equation (\ref{3.9}) implies
that the gravity wave amplitude can be comparable to $A_S$ if $\e$ is
sufficiently large. There exists a table of correspondences which
summarizes the four possibilities (Barrow \& Liddle 1993).

\begin{table}
\begin{center}
\begin{tabular}{c||c|c}
  Scalar      & Gravitational Waves        & Gravitational Waves     \\
  Spectrum    & Important                  & Negligible              \\
\hline
\hline
               &                           &                         \\
Small          & $e$ large                 & $e$ small               \\
Tilt           & $2\e \approx \eta$        & $|\eta|$ small          \\
               &                           &                         \\
Significant    & $\e$ large                & $\e$ small              \\
Tilt           & $|\eta|$ large            & $|\eta|$ large          \\
\end{tabular}
\end{center}
\footnotesize {\hspace*{.3in} Table 1 - The table of correspondences
illustrating the connection
between tilt and the magnitude of the energy and friction
parameters. The description `large' implies significantly larger than
zero (but still less than unity) and `small' implies the parameter is
very close to zero.}
\end{table}

\vspace{.35cm}

In the following section we shall employ this framework to derive the
form of the inflaton potential.

\vspace{.1cm}
\section{The  general potential leading to tilted spectra}
\vspace{.1cm}

Exact results may be derived by solving equation (\ref{3.7}). After
substitution of equation (\ref{3.6}) this becomes the first-order
differential equation
\be
\l{3.10}
\frac{dy}{dx} + \left( \frac{n-5}{2} \right) \frac{y}{x} + \left( \frac{1-n}{4}
\right) =0.
\ee
The solution for constant $n$ is
\be
\l{3.11}
y=\frac{1}{2} \left( \frac{1-n}{3-n} \right) x +C x^{(5-n)/2}
\ee
where $C$ is an arbitrary integration constant. The case $C<0$ is of
interest here because it leads to the {\em general} solution
\be
\l{3.12}
H(\phi) = \lambda \left[ {\rm sech} \left( \sqrt{
\frac{(n-1)(n-3)}{8}} \phi \right) \right]^{2/(3-n)}, \qquad n<1,
\ee
where
\be
\l{3.13}
\lambda = \left[ \frac{1}{2|C|} \frac{n-1}{n-3} \right]^{1/(3-n)}
\ee
Without loss of generality the integration constant has been removed
by performing a linear translation on the value of $\phi$. The
constant $|C|$ determines the energy scale at which these processes
are occurring (Carr \& Lidsey 1993; Lidsey \& Tavakol 1993). From
equation (\ref{10}) the potential is
\be
\l{potential}
V(\phi) =  \lambda^2 \left[ {\rm sech} (\omega\phi) \right]^{4/(3-n)} \left[ 3-
\left( \frac{n-1}{n-3} \right) {\rm tanh}^2(\omega\phi) \right] ,
\ee
where
\be
\l{potential1}
\omega \equiv \sqrt{ \frac{(n-1)(n-3)}{8}}.
\ee

The solution to  equation (\ref{3.11}) when $C=0$ is the exponential potential
\be
\l{3.14}
H(\phi) \propto \exp \left[ \pm \sqrt{\frac{1}{2} \left( \frac{n-1}{n-3}
\right)}\phi \right].
\ee
Eqs.  (\ref{3.12}) and (\ref{3.14}) lead to the same scalar
perturbation spectrum and a determination of $A_S(k)$ alone will not
lift the degeneracy. What distinguishes the two solutions is the
relative contribution of the gravitational wave spectrum, as
determined by equation (\ref{3.9}). We find
\be
\l{3.15}
\e = \left( \frac{n-1}{n-3} \right)
\ee
and
\be
\l{3.16}
\e = \left( \frac{n-1}{n-3} \right) \left[ {\rm tanh} \left( \sqrt{
\frac{(n-1)(n-3)}{8}}\phi \right) \right]^2
\ee
for the exponential and hyperbolic cases respectively. If cosmological
scales crossed the Hubble radius when $|\phi| \ll 1$, the amplitude of
tensor perturbations is exponentially suppressed in the latter
example.

Since equation (\ref{3.12}) is an exact solution, it is valid for all
values of $\phi$. In particular, for sufficiently small $\phi$ the
Taylor expansion
\be
\l{3.20}
H(\phi) = \lambda \left[ 1 - \left( \frac{1-n}{8} \right) \phi^2 +
{\cal{O}}(\phi^4) \right] \approx \sqrt{\frac{V(\phi)}{3}}
\ee
will also lead to a constant spectral index. This implies that a power
spectrum with $n={\rm constant} <1$ will arise from any function of
$H(\phi)$ which is identical to equation (\ref{3.20}) in the small
$\phi$ approximation.  It is well known that deviations from scale
invariance, without significant gravitational wave production, are
possible whenever the potential resembles an inverted harmonic
oscillator (Steinhardt \& Turner 1984), but the above calculation
provides further insight. Such a result follows because the inverted
oscillator resembles the hyperbolic secant function to lowest order in
$\phi$. Equation (\ref{3.20}) is very useful because it directly
relates the effective imaginary mass of the field to the scalar
spectral index.

These results are summarized pictorially in Figs. 1a and 1b, which are
representations of the class of solutions (\ref{3.11}) in the $x-y$
plane. ($x$ and $y$ are defined in equation (\ref{3.6})). In Fig. 1a
the coasting solution $y=x/2$ is shown as the dashed line and the
strong energy condition is violated to the right of this line. The
$x$-axis represents the de Sitter solution, $H={\rm constant}$, and
the origin is Minkowski space, which itself may be viewed as de Sitter
space with an infinite radius of curvature (Hawking \& Ellis
1973). The solid lines represent solutions of constant $n$ when
$C=0$. In these models $x$ is a measure of the energy density of the
universe and decreases as time increases. The trajectories of these
constant $n$ universes are indicated by the arrows and they all
asymptotically approach Minkowski space at $t \rightarrow
+\infty$.{\footnote{In reality the shape of the potential must change
at some point $(x,y)$ to allow for an exit from inflation.}}

\vspace{.3cm}
{\centerline{\bf Figure 1}}
\vspace{.3cm}

Fig. 1b illustrates the trajectories for finite values of $C<0$ and
$n=0.7$. This class of universe begins in a de Sitter phase at
$t=-\infty$ and evolves towards the $C=0$ asymptote at $t=+\infty$ in
such a way that the scalar spectral index remains constant at all
times. The magnitude of $C$ determines the amplitude of the scalar
quantum fluctuations but {\em not} the scale dependence.

Having found the form of the potential required, it is now necessary
to consider the physics which may lead to such a model.

\vspace{.3cm}
\section{Modelling the potential as a  bulk viscous stress}
\vspace{.3cm}

In this section we employ the techniques discussed in Section 2 for an
equation of state which has a number of physical applications in the
early universe.  Equation (\ref{3}) may be rewritten in the form $p
\equiv \G (\rho) \rho$ for some arbitrary function $\G$. The simplest
form for $\G (\rho)$ is the baryotropic equation of state
\be
\l{a1}
\G \equiv \gamma -1 = {\rm constant}, \qquad 0 \le \gamma \le 2,
\ee
which is equivalent at the classical level to the subset of solutions
\be
\l{a2}
{\dot{\phi}}^2=\mu V, \qquad \mu \equiv \frac{2\gamma}{2-\gamma}.
\ee
Hence $0 \le \mu \le \infty$, where $\mu =0$ $(\gamma =0)$ corresponds
to vacuum energy and $\mu = \infty$ $(\gamma =2)$ represents a stiff
fluid. In particular, radiation and matter dominated universes are
characterized by $\mu =4$ $(\gamma =4/3)$ and $\mu = 2$ $(\gamma =1)$
respectively. Furthermore, the equation of state for a universe
dominated by topological defects such as domain walls or a gas of
cosmic strings is also given by equation (\ref{a1}). It can be shown
that $\gamma_W=v^2_W+1/3$ for a gas of domain walls moving with
constant velocity $v_W$, whereas $\gamma_S=2(1+v_S^2)/3$ for a string
gas (Kolb \& Turner 1990).

Mathematically, a natural and plausible extension towards a more
realistic effective equation of state is to view the function $\G$ as
a truncated Taylor series
\be
\l{11}
\G (\rho) = (\gamma -1) - \beta \rho^{\alpha},
\ee
for positive definite constants $\{\alpha, \beta\}$. The second term
in this expression can be treated as a first-order perturbative
correction to $\G = {\rm constant}$. Physically, when $k=0$, equation
(\ref{11}) is equivalent to a fluid with bulk viscosity $\eta (\rho)
\equiv \Pi \rho^m$ and a perfect fluid $p=(\gamma -1)\rho$, where
$\beta = \sqrt{3} \Pi$ and $ m= \alpha +\frac{1}{2}$ (Barrow 1988). It
is known that the effects of particle production in the early universe
can be modelled in terms of a classical bulk viscosity of this
form. The polarization (trace anomaly) of quantized fields in curved
spacetimes also leads to vacuum viscosity effects (Zel'dovich 1980; Hu
1982; Waga, Falc\~ao \& Chanda 1986). Exact cosmological solutions
based on equation (\ref{11}) have been found by Murphy (1973) and
Barrow (1988), but in this section we derive them in a much simpler
parametric form. This allows the primordial fluctuation spectra to be
calculated.

With this choice of $\G(\rho)$ the Bianchi identity (\ref{2}) integrates
exactly for all $k$ to yield
\be
\l{12}
\rho(a)= \left[ \frac{\beta}{\gamma} +\left( \frac{a}{a_0}
\right)^{3\alpha\gamma} \right]^{-1/\alpha},
\ee
where the constant of integration is expressed in terms of $a_0$ and
$\{\alpha, \gamma \} \ne 0$. When $\beta =0$, $\rho \propto
a^{-3\gamma}$ as expected. Unfortunately analytical solutions to the
Friedmann equation (\ref{7}) have not been found when $k= \pm 1$, but
this equation simplifies to
\be
\frac{(\rho')^2}{\rho^2} = 3  \left( \gamma - \beta \rho^{\alpha} \right)
\ee
when $k=0$ and this has the exact solution
\bea
\l{13a}
1-\frac{\beta}{\gamma} \rho^{\alpha}(\phi) = {\rm tanh}^2 ( \omega\phi) \\
\l{13b}
\omega \equiv \sqrt{\frac{3\gamma\alpha^2}{4}}.
\eea
 The dominant energy condition is violated when $\beta <0$, so this
case is not considered further. The expressions for $H(\phi)$,
$a(\phi)$ and $V(\phi)$ follow from equations (\ref{9}) and (\ref{10})
as
\bea
\l{14a}
H(\phi) = \frac{1}{\sqrt{3}} \left( \frac{\gamma}{\beta} \right)^{1/2\alpha}
[{\rm sech}( \omega\phi)]^{1/\alpha} \\
\l{14b}
a(\phi) =a_0 \left( \frac{\beta}{\gamma} \right)^{1/3\alpha\gamma} \left| \sinh
(\omega\phi) \right|^{2/3\alpha\gamma} \\
\l{14c}
V(\phi) = \left( \frac{\gamma}{\beta} \right)^{1/\alpha} [ {\rm sech}
(\omega\phi) ]^{2/\alpha} \left( 1- ({\gamma}/2) {\rm tanh}^2(\omega\phi)
\right).
\eea
 It should be emphasized that these solutions are exact and no
`slow-roll' approximations, such as $|\ddot{\phi}| \ll H|\dot{\phi}|$
and ${\dot{\phi}}^2 \ll V$, have been made. These parametric solutions
are plotted schematically in Fig. 2.

\vspace{.3cm}
{\centerline{\bf Figure 2}}
\vspace{.3cm}

For completeness, we include the well known solutions for $\beta =0$, which
lead to the exponential potential
\be
\l{15a}
H(\phi) \propto \exp \left( \sqrt{\frac{3\gamma}{4}} \phi \right), \qquad
V(\phi) \propto \exp \left( \sqrt{3\gamma} \phi \right)
\ee
and power law expansion $a(t) \propto t^{2/3 \gamma}$.

Hence, when $k=0$, a perfect baryotropic fluid with bulk viscosity can
be modelled as a self-interacting scalar field with potential
(\ref{14c}). The inclusion of bulk viscosity alters the structure of
the potential away from an exponential form in the neighbourhood of
$|\phi| \approx 0$. This is illustrated in Fig. 2c. Near the origin,
we find that the equation of state (\ref{11}) can be adequately
described as an inverted harmonic oscillator. The exponential
potentials are recovered in the asymptotic limit as $|\phi|
\rightarrow \infty$, because the viscous effects decay faster than the
perfect fluid contribution at large $|\phi|$. The general advantage of
rewriting equation (\ref{3}) in terms of a scalar field is that the
qualitative history of the universe is easily determined by
considering the evolution of the scalar field along its interaction
potential. Here the field is initially placed at $\phi =0$, which
corresponds to a de Sitter expansion with $H=(\gamma /
\beta)^{1/2\alpha} / \sqrt{3}$. As the field rolls either to the left
or right of the potential, the universe expands and approaches the
power law attractor solution $a \propto t^{2/3\gamma}$ with $H
\rightarrow 0$ monotonically. This behaviour was first noted by Barrow
(1988).

The region of parameter space in which the strong energy condition
$(\rho +3p >0)$ is violated can be determined from equations
(\ref{11}) and (\ref{13a}). Defining the quantity $\Delta \equiv \rho
+3p$, we find
\be
\l{15}
\Delta =3\gamma\rho \left( {\rm tanh}^2(\omega\phi) -\frac{2}{3\gamma} \right).
\ee
It follows that inflation occurs for all $\phi$ when $\gamma < 2/3$
and there exists a graceful exit problem. An identical problem is
encountered when $\beta =0$. The potential must be modified in such a
way that allows the expansion to become subluminal. For $\gamma >
2/3$, on the other hand, the universe inflates initially, but deflates
once
\be
\l{16}
{\rm tanh}^2{\omega\phi} > \frac{2}{3\gamma}
\ee
is satisfied. The end of inflation can be defined as the point where this
becomes an equality.

The epoch during inflation which is relevant for large-scale structure
observations can then be determined and we now proceed to discuss
models which lead to a potential of the form given by equation
(\ref{11}).

\vspace{.3cm}
\section{Particle physics models}

\vspace{.3cm}

The purpose of this section is to discuss a number of particle physics
models in which the features described above may arise.

\vspace{.2cm}
{\centerline{\bf A. Bulk Viscosity}}
\vspace{.2cm}

A direct comparison of Eqs. (\ref{3.12}) and (\ref{14a}) indicates
that tilted spectra with constant spectral index and negligible
gravitational waves arise in a class of bulk viscosity models if we
identify
\be
\l{5.1}
\alpha = \frac{3-n}{2}
\ee
\be
\l{5.2}
\gamma = \frac{2}{3} \left( \frac{n-1}{n-3} \right) \qquad \Longleftrightarrow
\qquad n = \frac{9\gamma -2}{3\gamma -2},
\ee
or equivalently
\be
\l{5.3}
\alpha = \frac{2}{2-3\gamma}.
\ee
The constraint $n \ge 0.7$ now becomes the upper limit $\gamma \le
0.09$. (We require $\gamma < 2/3$ for consistency in this class, since
equation (\ref{14a}) is only valid for positive $\alpha$). The
parameter $\beta$ of equation (\ref{11}) plays the role of $|C|$ in
determining the amplitude of the fluctuations, whereas $\alpha$
determines the tilt of the spectrum.

When $\{\alpha, \gamma \}$ are not related by equation (\ref{5.3}) the
tilt is not exactly scale invariant, but $n \approx {\rm constant}$ is
an excellent approximation if $\omega |\phi| \ll 1$. For standard
reheating, scales of astrophysical interest first crossed the Hubble
radius approximately $50$ e-foldings before the end of
inflation. Defining the value of the field at this point as
$\phi_{50}$, we find from equation (\ref{16}) that
\be
\l{5.4}
{\rm sinh}^2(\omega\phi_{50}) = \left( \frac{2}{3\gamma -2} \right)
e^{-3\alpha\gamma N_{50}}
\ee
for $\gamma >2/3$, where $N_{50} \approx 50$. Hence, if $\gamma $ is
not too close to $\gamma = 2/3$, $\omega\phi_{50} \ll 1$ is valid and
it is consistent to expand equation (\ref{14a}) to lowest order. It
follows from equation (\ref{3.20}) that
\be
\l{5.5}
n=1-3\alpha\gamma
\ee
and $n\ge 0.7$ leads to the constraint
\be
\l{5.6}
\alpha\gamma \le 0.1.
\ee

\vspace{.2cm}
{\centerline{\bf B. Quantum creation of fundamental strings }}
\vspace{.2cm}

Turok (1988) has considered the quantum production of infinitely thin
Witten strings on super-horizon size scales (Green, Schwarz \& Witten
1988). He suggested that a deflationary expansion follows naturally
from a quasi de Sitter phase in the early universe. When $\alpha =1$
and $2/3 \le \gamma \le 1$, it was shown that the equation of state
(\ref{11}) provides a good phenomenological description of the quantum
creation of these strings after compactification to four dimensions
(Barrow 1988; Turok 1988). The parameter $\beta$ depends on the
fundamental string tension and the fractal dimension of the string.
The lower and upper limits on $\gamma$ correspond to long strings with
negligible velocity and a highly convoluted, relativistic string
distribution respectively.

Therefore, the parametric solutions (\ref{14a})-(\ref{14c}) describe
the evolution of the flat Friedmann universe when dominated by
fundamental strings created on super-horizon scales.  It follows from
equation (\ref{5.5}) that $-2 \le n \le -1$ and the model in its
present form gives far too much power on large scales. However, it is
interesting that this predicted range for the spectral index
corresponds precisely to that observed on intermediate scales $1 \le
\lambda \le 200 h^{-1}$ Mpc in the distribution of clusters of
galaxies (Einasto et al. 1993).

\vspace{.2cm}
{\centerline{\bf C. $N=2$ supergravity in six dimensions}}
\vspace{.2cm}

The Salam-Sezgin model is $N=2$ supergravity in six dimensions
compactified onto a two-sphere (Salam \& Sezgin 1984). Although such a
model does not reproduce a correct particle spectrum in four
dimensions, it is thought to contain features generic to more complete
theories and its cosmological implications have been studied in some
detail (Liddle 1989). It has been shown that the theory is equivalent
to two interacting scalar fields, $\xi$ and $\sigma$, whose full
potential exhibits a global minimum in both $\xi$ and $\sigma$
directions if the scale-invariance of the theory is broken by quantum
effects (Gibbons \& Townsend 1987). If the $\xi$-field comes to rest,
the potential for $\sigma$ becomes
\be
\l{5.7}
V(\sigma) =V_0 \left[ \exp (-\sqrt{8}\sigma) -2\exp (-\sqrt{2}\sigma ) +1
\right].
\ee
The potential has a global minimum located at $\sigma=0$ and an
asymptotic form $V \propto \exp (-\sqrt{8}\sigma)$ for $\sigma \ll
0$. In this regime the system behaves as a baryotropic perfect fluid
with $\gamma = 8/3$. In this unmodified form the model does not lead
to inflation, but it is clear from Sections 3 and 4 that introducing a
bulk viscosity into the system will violate the strong energy
condition and lead to a tilted spectrum. By condition (\ref{5.6})
observation requires $\alpha < 0.038$.

Moreover, the viscosity effects redshift faster than the perfect fluid
contribution, implying that they will become negligible by the time
the field has rolled towards $\sigma =0$. Hence, the shape of the
potential is not significantly altered within the vicinity of the
global minimum and standard reheating can occur through rapid
oscillations of the field about this minimum (Kolb \& Turner
1990). This solves the graceful exit problem discussed in Section 2.

\vspace{.2cm}
{\centerline{\bf D. Natural Inflation}
\vspace{.2cm}

For completeness we remark that natural inflation, driven by a
pseudo-Nambu-Goldstone boson with potential $V(\phi) \propto 1 + \cos
(\phi /f)$, also produces a tilted spectrum. This model has been
discussed in detail by Adams et al. (1993). In the small angle
approximation the spectral index is given by $n = 1 -f^{-2}$ via
equation (\ref{3.20}).

\vspace{.3cm}
\section{Conclusions}
\vspace{.3cm}

The general inflationary potential which produces a measureable tilt
in the primordial spectrum of scalar fluctuations, without producing a
significant gravitational wave spectrum, is of the form
(\ref{potential}).  In this paper it has been shown that such a
potential is physically equivalent to a bulk viscous stress
modification to a baryotropic perfect fluid. Mathematically it is
equivalent to expanding the equation of state as a truncated Taylor
series. It was further shown that this potential arises in a number of
particle physics theories.  If future observations indicate that a)
gravitational waves are not contributing to the anisotropy in the
cosmic microwave background and b) a dark matter model based on a
spectrum with significant tilt does provide a good fit to the data,
then this would justify a more detailed study of the models discussed
in Section 5.

A number of simplifying assumptions were made. In particular, the
expressions (\ref{3.1}) and (\ref{3.2}) for the scalar and tensor
amplitudes are strictly only valid in the slow roll regime, $\{\e,
|\eta|\} \ll 1$, whereas equation (\ref{3.7}) implies that a tilt away
from the Harrison-Zel'dovich spectrum requires $ 0 \ll |\eta| \le 1 $,
approximately. However, it has been shown that the corrections away
from slow-roll are not important near a local maximum, and although
they slightly alter the amplitude of the fluctuations in the
exponential regime, they do not change the spectral index (Stewart \&
Lyth 1993).

The physical interpretation of the potential in terms of a bulk
viscosity is only valid in the spatially flat FRW cosmology. However,
this paper has investigated the power spectra of such models and it is
the last $60$ e-foldings of inflationary expansion which are important
for large-scale structure (Kolb \& Turner 1990). In most chaotic
scenarios the density parameter is very close to unity by this stage.

Furthermore, since all scales probed by large-scale structure
correspond to a small $( \approx 9)$ number of e-foldings, it is
reasonable to assume that the parameter $\gamma$ is constant during
this interval. These results may therefore have wider applications in
models where the polytropic index is a function of cosmic time.

Finally we note that equation (\ref{11}) with $\gamma =1$ and $\beta
<0$ is the equation of state for a polytropic star, special cases of
which include white dwarfs $(\alpha =5/3)$ and neutron stars $(\alpha
= 4/3)$ (Weinberg 1973). If the solution to equation (\ref{7}) could
be found for $k=+1$, the techniques described here could be relevant
for stellar structure.

\vspace{1in}
{\bf Acknowledgments} \hspace*{2em} The author is supported by the
Science and Engineering Research Council (SERC) UK and is supported at
Fermilab by the DOE and NASA under Grant NAGW-2381.

\newpage

\frenchspacing
\def\prl#1#2{{ Phys. Rev. Lett.,} {#1}, #2}
\def\prd#1#2{{ Phys. Rev. D,} {#1}, #2}
\def\plb#1#2{{ Phys. Lett. B,} {#1}, #2}
\def\npb#1#2{{ Nucl. Phys. B,} {#1}, #2}
\def\apj#1#2{{ ApJ,} {#1}, #2}
\def\apjl#1#2{{ ApJ Lett.,} {#1}, #2}

\vspace{1.0in}
\centerline{{\bf References}}
\begin{enumerate}

\item [] Abbott L. F., Wise M. B., 1984, \npb{244}{541}

\item []  Adams F. C.,   Bond J. R.,  Freese K.,   Frieman J. A.,   Olinto A.
V., 1993, \prd{47}{426}

\item []   Albrecht A.,   Steinhardt P. J., 1982, \prl{48}{1220}

\item []  Barrow J. D., 1988, \npb{310}{743}

\item []  Barrow J. D., 1990, \plb{235}{40}

\item []  Barrow J. D.,   Liddle A. R., 1993, `Perturbation Spectra from
Intermediate Inflation, Sussex preprint SUSSEX-AST 93/2-1

\item []   Bertschinger E., Dekel A., 1989, \apjl{336}{L5}

\item []    Bertschinger E., Dekel A.,  Faber S. M.,  Dressler A.,  Burstein
D., 1990, \apj{364}{370}

\item []   Birrell N. D.,   Davies P. C. W.,  1982, Quantum Fields in Curved
Space. Cambridge Univ. Press, Cambridge

\item [] Carr B. J., Lidsey J. E., 1993, Phys. Rev. D, to appear

\item []  Copeland E. J.,   Kolb E. W.,   Liddle A. R.,   Lidsey J. E., 1993,
`Reconstructing the Inflaton Potential - In Principle and in Practice',
Sussex/Fermilab preprint SUSSEX AST 93/3-1, FNAL-PUB-93/029-A, to appear  Phys.
Rev.  D; and refs therein

\item []   Davies M.,   Efstathiou G.,   Frenk C. S.,   White S. D. M., 1985,
\apj{292}{371}

\item []  Dekel A.,  Bertschinger E.,   Faber S. M., 1990, \apj{364}{349}

\item []  Efstathiou G., 1990,  in  Heavens A.,  Peacock J.,  Davies A., eds,
The Physics of the Early Universe. SUSSP publications

\item []   Efstathiou G.,  Rees M. J., 1988, MNRAS,   230 , 5P

\item []   Efstathiou G., et al.,  1990, MNRAS,  247 , 10P

\item []  Einasto J., Gramann M., Saar E., Tago E., 1993, MNRAS, 260, 705

\item []   Gibbons G. W., Townsend P. K., 1987, \npb{282}{610}

\item []   Green M. B.,  Schwarz J. H.,  Witten E., 1988,  Superstring Theory.
Cambridge Univ. Press, Cambridge

\item []   Grishchuk L. P., 1975,  Soviet Phys. JETP,  40, 409

\item []  Grishchuk L. P., 1977, Ann. N. Y. Acad. Sci.,   302, 439

\item []  Guth A. H., 1981, \prd{23}{347}

\item []  Hawking S. W., Ellis G. F. R.,   The Large Scale Structure of
Spacetime. Cambridge Univ. Press, Cambridge

\item []  Hu B. L., 1982, Phys. Lett. A,  90, 375

\item []  Kolb E. W.,  Turner M. S., 1990, The Early Universe. Addison-Wesley,
New York

\item []  Liddle A. R., 1989, \plb{220}{502}; and refs therein

\item [] Liddle A. R.,  Lyth D. H., 1993, `The Cold Dark Matter Density
Perturbation'.   To appear Phys. Repts

\item []  Lidsey J. E., 1991 \plb{273}{42}

\item [] Lidsey J. E., 1993,  Gen. Rel. Grav., 25, 399

\item []  Lidsey J. E., Tavakol R. K., 1993, `On the Correspondence between
Theory and Observation in Inflationary Cosmology', Fermilab preprint
FNAL-PUB-93/034-A (1993). To appear, Phys. Lett. B

\item [] Linde A. D., 1982, \plb{108}{389}

\item []  Maddox S. J.,   Efstathiou G., Sutherland W. J.,  Loveday J.,  1990,
MNRAS,  242, 43P

\item []  Maddox S. J.,  Sutherland W. J., Efstathiou G.,  Loveday J.,
Peterson B. A., 1991, MNRAS, 247, 1P

\item []  Murphy G. L.,  1973, \prd{8}{4231}

\item []  Muslimov A. G., 1990, Class. Quantum Grav., 7, 231

\item [] Olive K. A., 1990,  Phys. Repts., 190, 307

\item [] Peacock, J. A., 1991, MNRAS, 253, 1P

\item [] Raychaudhuri A. K., 1955, Phys. Rev., 98, 1123

\item [] Salam A.,   Sezgin E., 1984, \plb{147}{47}

\item []  Salopek  D. S., Bond J. R., 1990, \prd{28}{2960}

\item [] Salopek D. S.,  Bond J. R., 1991, \prd{43}{1005}

\item []  Saunders W.,  et al., 1991, Nat., 349, 32

\item [] Stewart  E. D.,  Lyth D. H., 1993, \plb{302}{171}

\item []  Smoot  G. F., et al., 1992, \apjl{396}{L1}

\item []  Steinhardt P. J.,  Turner M. S., 1984, \prd{29}{2162}

\item [] Turok N., 1988, \prl{60}{549}

\item []  Waga I.,  Falc\~ao R. C.,  Chanda R., 1986, \prd{33}{1839}

\item []  Weinberg S. W., 1973, Gravitation and Cosmology. Wiley and Sons, New
York

\item []  White M., 1992, \prd{46}{4198}

\item []  Wright E. L.,  et al., 1992, \apjl{396}{L13}

\item []  Zel'dovich Ya. B., 1980, Sov. Phys. JETP Lett.,  12, 307

\end{enumerate}

\newpage

{\centerline{\bf Figure Captions}}

\vspace{1cm}

{\em Figure 1a}: The trajectories of the constant $n$ universes in the
$x-y$ place when $C=0$. The dashed line represents the Milne universe
and the strong energy condition is satisfied to the left of this
line. All trajectories approach the origin at $t=+\infty$, which is de
Sitter space with infinite radius of curvature.

\vspace{1cm}

{\em Figure 1b}: The trajectories of $n=0.7$ universes when $C=\{0,
-0.5, -1, -2\}$. The dashed line corresponds to $C=0$. All solutions
begin in a de Sitter phase at $t=-\infty$ and approach the $C=0$
(power law) asymptote at $t=+\infty$.

\vspace{1cm}

{\em Figure 2}: Schematic plots of the parametric solution (38)-(40)
as a function of the scalar field. In Fig. 2c the dashed lines
represent the exponential potential corresponding to vanishing bulk
viscosity. The inclusion of a bulk viscous stress results in a local
maximum in the potential and a quasi de Sitter expansion. It is this
feature which results in negligible gravitational wave production, but
the curvature of the potential at the origin tilts the scalar
spectrum.

\end{document}